\documentclass[a4paper]{article}

\usepackage[english]{babel}
\usepackage[utf8x]{inputenc}
\usepackage[T1]{fontenc}

\usepackage[a4paper,top=3cm,bottom=2cm,left=3cm,right=3cm,marginparwidth=1.75cm]{geometry}

\usepackage{amsmath}
\usepackage{graphicx}
\usepackage[colorinlistoftodos]{todonotes}
\usepackage[colorlinks=true, allcolors=blue]{hyperref}
\usepackage{xspace}

\newcommand{\eg}{e.g.,\xspace}
\newcommand{\ie}{i.e.,\xspace}
\newcommand{\reproserver}{{\fontfamily{lmss}\selectfont ReproServer}\xspace}
\newcommand{\reprozip}{ReproZip\xspace}
\newcommand{\reprozipsoftware}{\texttt{reprozip}\xspace}
\newcommand{\reprounzipsoftware}{\texttt{reprounzip}\xspace}
\newcommand{\rpz}{\texttt{.rpz}\xspace}
\newcommand{\reprounzipdocker}{\texttt{reprounzip-docker}\xspace}
\newcommand{\reprounzipvagrant}{\texttt{reprounzip-vagrant}\xspace}

\renewcommand{\paragraph}[1]{\vspace{0.1cm}\noindent \textbf{#1}.}

\title{ReproServer: \\Making Reproducibility Easier and Less Intensive}
\author{}

\author{
  R\'{e}mi Rampin\\
  \texttt{remi.rampin@nyu.edu}
  \and
  Fernando Chirigati\\
  \texttt{fchirigati@nyu.edu}
  \and
  Vicky Steeves\\
  \texttt{vicky.steeves@nyu.edu}
  \and
  Juliana Freire\\
  \texttt{juliana.freire@nyu.edu} 
}

\date{}

\begin{document}
\maketitle
\section{Introduction}
The adoption of reproducible research methods in academia remains low, despite the altruistic and self-centered reasons for supporting reproducibility individually and at large. The more magnanimous are the ideas that reproducibility aids in driving a research field forward, inducting newcomers to both standard and cutting-edge methods, teaching, verifying existing work, and reusing previously published research. 

Currently, researchers have to rely on tables, figures, and plots included in papers to get an idea of the research results, which notably hide the details on how they were derived. As the methods and workflows can be very complex and dependent on multiple parameters that are hardly described in papers, reproducing the results from scratch is barely feasible, if not impossible.

Reproducibility may be difficult to achieve for a number of reasons; in particular, it can be technically challenging. When working with different software and scripts, these may have many nested library dependencies, of which different versions may give entirely disparate results or not even run~\cite{gronenschild}. To address this problem, colloquially known as \textit{dependency hell}, researchers ideally have to create a compendium that encapsulates all the dependencies needed for reproducibility, including data, code, and information about the computational environment (\eg operating system, hard/software architecture, and library dependencies). However, manually keeping track of this information is rarely feasible for many reasons. Computational environments are complex, consisting of many layers of software, and the configuration of the operating system is often hidden. In addition, tracking software library dependencies is exceptionally challenging, especially for large experiments. Even if such compendium is available, it may be still hard to reproduce the results. There may be no instructions about how to execute the code and explore it further; the code may not run on a different operating system; there may be missing libraries; library versions may be different; and issues may arise while trying to install all the required dependencies.

The aforementioned technical challenge is a problem across many facets of research, \eg when working in a collaboration, when attempting to build upon previous work, and when training new graduate students in a lab. In particular, this is an issue faced when researchers undergo peer review of their work. Reviewers may not have access to the corresponding code and data, and thus, have to solely rely on the textual descriptions. Even when they have access to the computational assets, it may not only be hard but also \textbf{\textit{intrusive}} to run the code, since they have to install dependencies in their own machine. While reviewers can make use of a virtual machine to isolate the review from their own environment, the entire process is still significantly \textbf{\textit{time-consuming}}. These barriers end up further discouraging reproducible research, which still remains an elusive target for many researchers.

There are solutions that have worked in other fields that don't necessarily work for research applications. Continuous Integration (CI) is used to ensure software works reproducibly, in that it allows developers to maintain a shared code repository, and institute an automated build that includes testing scripts to ensure everything works before deploying. CI is very common and free to use (e.g. GitLab CI and Travis CI) and some even go so far as to allow developers to choose a container with a particular computational environment to build and deploy software, even allowing each job to run in a separate and isolated container. But no such solution exists for science, where researchers have to test their analyses locally. There are a few platforms which researchers can use to make their work somewhat reproducible, but these are costly and do not allow users to export their work in a reproducible way out of the platform. There are other open source and free platforms that work only with specific tools, such as Binder for reproducing Jupyter Notebooks.

While some solutions have been introduced recently, they only partially solve the problem. For instance, Binder~\cite{binder} can reproduce interactive notebooks directly from an URL or a GitHub repository, allowing users to interact with these notebooks in a live environment from a Web browser. However, users must provide a file that describes the dependencies for the notebooks (\eg a \texttt{requirements.txt} file or a Dockerfile), \ie users must manually capture such dependencies.
Code Ocean~\cite{codeocean} is a proprietary, closed-source platform for reproducible research that allows users to host code and data on the Web. Others can then view, edit, and run the code using the Web browser and share it privately or publicly. Nevertheless, users still need to manually capture all the files and dependencies to add to their servers. Also, this is a costly solution and does not allow users to export their work in a reproducible way \textit{out of the platform}.

\section{Our Previous Work: \reprozip}

Recently, we have introduced \reprozip~\cite{chirigati@sigmod2016,joss}, an open-source desktop application that alleviates part of the aforementioned problems. \reprozip allows users to automatically and transparently capture all the dependencies of a computational experiment in a single, distributable bundle, that can be used to reproduce the entire experiment in another environment. Users do not need to start their project with reproducibility in mind: \reprozip works automatically given an \textit{existing} application, independently of the programming language. Overall, the tool works in two steps:

\begin{itemize}
\item \textbf{Packing.} Users can run their experiment with the \reprozipsoftware component to automatically capture it in totality, including software, data files, databases, libraries, environment variables, parameters, and OS and hardware information. \reprozipsoftware then creates a compendium of their work---a \rpz file---that is easily shareable, citable (if made public in a repository), and usable by themselves and the community at large.
\item \textbf{Unpacking.} Given a \rpz file, other users can use the \reprounzipsoftware component of \reprozip to set up the experiment in their environment, even if their operating system is different than the original one. Users can choose their unpacker of choice (\eg Vagrant or Docker for an isolated reproduction), and \reprounzipsoftware automatically sets up the environment. \reprounzipsoftware has also many user-friendly interfaces to make the reproduction even easier, including a graphical user interface, and a way to replace input files and parameters for testing the behavior of the experiment under different inputs.
\end{itemize}

\reprozip has been successfully used to reproduce myriad experiments, including Python scripts, R scripts, client-server applications (that include a database), and GUI applications; please refer to~\cite{reprozip-examples} for examples. \reprozip has also been recommended by different venues, including the Information Systems Journal~\cite{information-systems}, ACM SIGMOD~\cite{sigmod}, and conferences that follow the Artifact Evaluation Process guidelines~\cite{aec}.

Note that \reprozip bridges two gaps: it \textit{automatically} captures the dependencies of an experiment, and it \textit{automatically} configures these dependencies in a new environment. Therefore, it makes it easier for authors to make their research reproducible, and reviewers to run the application associated with the research. Because the reproduction can be performed in an isolated manner (using Vagrant or Docker), it is not intrusive either.

While \reprozip significantly reduces the barrier to reproducibility, to unpack and reproduce \rpz bundles, users must still download the \reprounzipsoftware component and their unpacker of choice (\reprounzipdocker to use Docker, or \reprounzipvagrant to use Vagrant). In addition, they also need to install the software to be used by the unpacker (\ie Docker, or Vagrant and VirtualBox). Even if this is only required once, having to download and set up these tools can prove to be a heavy burden (and intrusive), especially in the author-reviewer scenario where reviewers have a short turnaround time. Therefore, we seek a platform that mitigates the time necessary for reviewers to reproduce other people's work, expediting the review process.

\section{Our Solution: ReproServer}

In this report, we present \reproserver~\cite{reproserver}, an open source Web application that allows users to reproduce experiments from the comfort of their Web browser. \reproserver leverages ReproZip: users can unpack and interact with \reprozip bundles over the Web, \textit{without having to download any software}, and finally \textit{share persistent links} to the unpacked versions. The user experience is, therefore, similar to that of \reprounzipsoftware, but with the following additional advantages: (1) everything is done from the Web, and (2) users can easily share the reproduction environment.

\begin{figure}[t]
\centering
\includegraphics[width=1.0\textwidth]{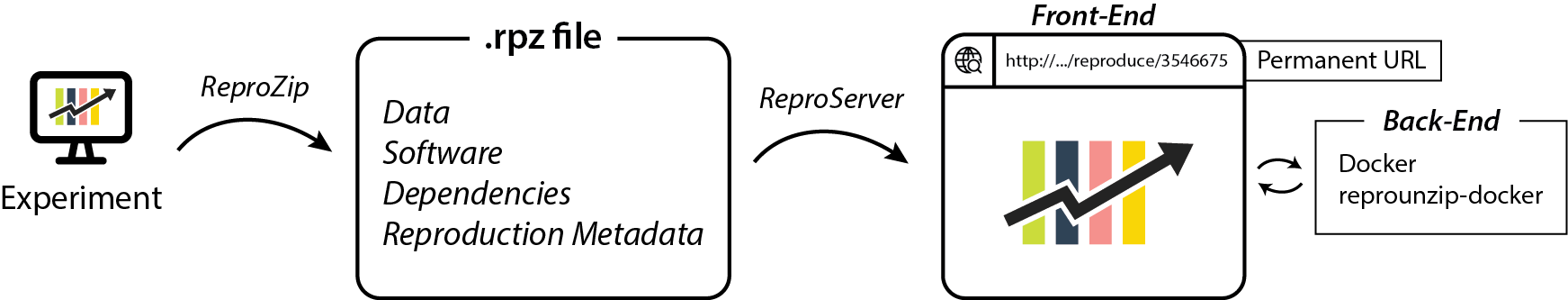}
\caption{Overall flow of \reproserver.}
\label{fig:flow}
\end{figure}

To use \reproserver, one simply needs to either \textit{upload a \rpz file from their machine} or \textit{provide a link to a \rpz file}. Upon successful upload of the \rpz file, users can interact with the experiment before rerunning it. In this interface, users can tweak parameters and even upload new inputs, which is particularly useful for testing the experiment against similar datasets. Then, in one click, they can re-execute the contents of the \rpz file, explore the log, download the output files, and thus verify and build on other people's work easily, without having to install any software. The overall flow is depicted in Figure~\ref{fig:flow}.

\reproserver has also support for bundles stored in the Open Science Framework~\cite{osf} and figshare~\cite{figshare}. In this case, the identifier for a bundle can be included directly into the \reproserver link: in one click, others can reproduce the \rpz bundle (hosted on a different service) via \reproserver, which provides stronger \textit{persistence} guarantees. For instance, if a \rpz bundle has identifier \texttt{3546675} on figshare, anyone can reproduce this bundle by using the path \texttt{/reproduce/figshare.com/3546675} under \reproserver. If the bundle is stored elsewhere or is uploaded, \reproserver also provides a short identifier. With such mechanism, \reproserver is able to generate permanent URLs that are easy to include in publications, making it easier to provide \textit{one-click reproduction}.

\subsubsection*{Example}

Alice is the author of a paper to be reviewed soon, and she wishes to make the corresponding experiment fully open and reproducible. The experiment has three Python scripts: one for data collection (\texttt{collection.py}), one for data analysis (\texttt{analysis.py}), and one for plotting the results (\texttt{plots.py}). Alice first uses \reprozip to create a \rpz bundle for her experiment. Instead of running the scripts in the regular way

\vspace{.3cm}
{\ttfamily
\noindent python collection.py \\
python analysis.py \\
python plots.py
}
\vspace{.3cm}

\noindent she prepends \reprozipsoftware to the execution:

\vspace{.3cm}
{\ttfamily
\noindent reprozip trace python collection.py \\
reprozip trace -{}-continue python analysis.py \\
reprozip trace -{}-continue python plots.py
}
\vspace{.3cm}

\noindent When her scripts finish running, she creates the \reprozip bundle using the following:

\vspace{.3cm}
{\ttfamily
\noindent reprozip pack experiment.rpz
}
\vspace{.3cm}

\noindent This generates the \rpz file \texttt{experiment.rpz}, which contains all the dependencies necessary to reproduce her work.

Alice then posts her \rpz file on the Open Science Framework and uses the link to the file in \reproserver to test whether it successfully reruns. After rerunning the scripts and successfully reproducing the experiment, \reproserver generates a \textit{link} to the unpacked environment, and Alice includes this link in her paper, inviting others to reproduce and extend their work.

During the review process, one of the reviewers, named Bob, wishes to verify the results that Alice discusses in her paper. Bob clicks on the \reproserver link that Alice added to the paper. The link takes Bob to his Web browser, where \reproserver opens the unpacked environment. In one click, he is able to fully re-execute and interact with the original experiment, without having to install any software.

\subsection{Architecture}

\reproserver has two key components: the \textit{front-end}, where users interact with the system through their preferred Web browser, and the \textit{back-end}, which builds the environment from the metadata included in the original \reprozip bundle and re-executes the experiment in that environment.

\subsubsection{Front-End}

Users interact with the front-end via an initial interface, which asks for a local \reprozip bundle or a link to a \reprozip bundle. Once the \rpz file is successfully uploaded, users are exposed to the original experiment parameters and input files, which they can modify if desired (for instance, to test how consistent the experiment is under different inputs). Users then click the \textit{Run} button and \reproserver executes the contents of the \rpz file. After the re-execution, \reproserver places the output files on an Amazon S3-compatible storage service for users to download. During and after the re-execution, \reproserver also shows the execution log.

This process generates for the user a permanent URL to share with people and embed in articles, allowing others to reproduce the experiment. Currently, \reproserver has support for the Open Science Framework and other data repositories for persistence of data (including figshare), instead of relying on \reproserver for storage. Using one of those repositories also allow for greater discoverability of work. The main goal with this approach is to integrate as much as possible into researchers' existing workflows. For direct uploads, this URL has a ``local'' short identifier without persistence guarantees; for repositories, \reproserver uses repository-native identifiers (\eg \texttt{/reproduce/osf.io/5ztp2}, \texttt{/reproduce/figshare.com/3546675}). Upon clicking on the permanent URL, it takes users to an interface to set parameters and upload new input data (if desired) for the reproduction.

\subsubsection{Back-End}

From the \rpz file, \reproserver builds a Docker image and caches it on a private Docker registry for later execution. If the file is hosted on a data repository, \reproserver can rebuild this image at any time. However, caching the images has the advantage of speeding up the reproduction and catching malformed files earlier in the process. \reproserver also extracts the metadata from the bundle into the database at this time, \eg the list of input and output files and the original command lines.
\begin{figure}
\centering
\includegraphics[width=1.0\textwidth]{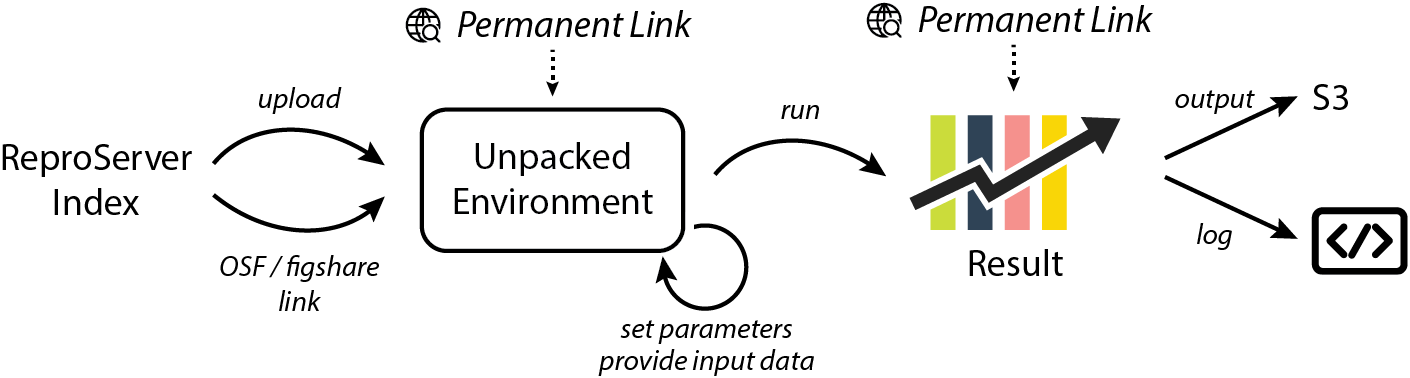}
\caption{\label{fig:commands}ReproServer commands.}
\end{figure}

From the cached Docker image, input files, and parameters, \reproserver runs the experiment and stores the results. We enforce a run time and memory limit to prevent abuse and keep the service accessible to all. More machines can be provisioned to run experiments in response to traffic using Kubernetes' auto-scaling feature.

\subsubsection{Deployment}

Easy setup scripts are included, which allow for local testing and development, as well as deployment of the system for production on a Kubernetes cluster (for example, on Google Cloud Platform). Many aspects of the software can be configured via a configuration file, which also allows for multiple deployments (for staging new developments) and rolling updates.

\section{Conclusion}

In this report, we introduced \reproserver, a Web application that allows users to easily reproduce experiments in one click. We envision \reproserver, together with \reprozip, as a framework for easy reproduction without being intrusive: research can be easily and automatically packed on one side in very few steps, and they can be automatically unpacked and reproduced on the other side without having to install additional software. As we increasingly lower the barrier for reproducibility, we hope that more research is made reproducible in the future and less excuses are given for not making code and data openly available, thus truly moving science forward.

\bibliographystyle{abbrv}
\bibliography{main}

\begin{thebibliography}{10}

\bibitem{sigmod}
{ACM SIGMOD Reproducibility Review}.
\newblock \url{http://db-reproducibility.seas.harvard.edu/}.
\newblock Last Access: Nov 2017.

\bibitem{aec}
{Artifact Evaluation Process Guidelines}.
\newblock \url{http://www.artifact-eval.org/guidelines.html}.
\newblock Last Access: Nov 2017.

\bibitem{binder}
{Binder}.
\newblock \url{https://mybinder.org/}.
\newblock Last Access: Nov 2017.

\bibitem{codeocean}
{Code Ocean}.
\newblock \url{https://codeocean.com/}.
\newblock Last Access: Nov 2017.

\bibitem{figshare}
{figshare}.
\newblock \url{https://figshare.com/}.
\newblock Last Access: Nov 2017.

\bibitem{information-systems}
{Information Systems Journal, Elsevier}.
\newblock \url{https://www.journals.elsevier.com/information-systems/}.
\newblock Last Access: Nov 2017.

\bibitem{osf}
{Open Science Framework (OSF)}.
\newblock \url{https://osf.io/}.
\newblock Last Access: Nov 2017.

\bibitem{reproserver}
{ReproServer}.
\newblock \url{https://github.com/ViDA-NYU/reproserver}.
\newblock Last Access: 2017.

\bibitem{reprozip-examples}
{ReproZip Examples}.
\newblock \url{https://examples.reprozip.org//}.
\newblock Last Access: Nov 2017.

\bibitem{chirigati@sigmod2016}
F.~Chirigati, R.~Rampin, D.~Shasha, and J.~Freire.
\newblock {ReproZip: Computational Reproducibility With Ease}.
\newblock In {\em Proceedings of the 2016 International Conference on
  Management of Data}, SIGMOD '16, pages 2085--2088, New York, NY, USA, 2016.
  ACM.

\bibitem{gronenschild}
E.~H. B.~M. Gronenschild, P.~Habets, H.~I.~L. Jacobs, R.~Mengelers,
  N.~Rozendaal, J.~van Os, and M.~Marcelis.
\newblock {The Effects of FreeSurfer Version, Workstation Type, and Macintosh
  Operating System Version on Anatomical Volume and Cortical Thickness
  Measurements}.
\newblock {\em PLOS ONE}, 7(6):1--13, 06 2012.

\bibitem{joss}
R.~Rampin, F.~Chirigati, D.~Shasha, J.~Freire, and V.~Steeves.
\newblock {ReproZip: The Reproducibility Packer}.
\newblock {\em The Journal of Open Source Software}, 1(8), Dec 2016.

\end{thebibliography}

\end{document}